\documentclass[preprint2]{aastex}
\usepackage{psfig}
\newcommand{\dopp}{$\mathcal D$}
\begin{document}
 
\title{The discovery of a microarcsecond quasar: J\,1819+3845}
\shorttitle{The discovery of a microarcsecond quasar}

\author{J. Dennett-Thorpe\altaffilmark{1,3}, A.G. de Bruyn\altaffilmark{2,1}}
\altaffiltext{1}{Kapteyn Institute, Rijksuniversiteit Groningen, 9700
AV Groningen, The Netherlands}
\altaffiltext{2}{Netherlands Foundation for Research in Astronomy,
7990 AA Dwingeloo, The Netherlands}
\altaffiltext{3}{Observatorio Astronomico, Lisbon, Portugal}
\email{jdt@astro.rug.nl}

\begin{abstract}
 We report on the discovery of a source which exhibits over 300\%
amplitude changes in radio flux density on the period of hours. This
source, J\,1819+3845, is the most extremely variable extragalactic
source known in the radio sky. We believe these properties are due to
interstellar scintillation, and show that the source must emit at
least 55\% of its flux density within a radius of $<$16
microarcseconds at 5\,GHz. The apparent brightness temperature is
$>$\,5.10$^{12}$ K, and the source may be explained by a
relativistically moving source with a Doppler factor $\sim$ 15. The
scattering occurs predominantly in material only a few tens of parsecs
from the earth, which explains its unusually rapid variability. If the
source PKS\,0405-385 (Kedziora-Chudczer et al 1997) is similarly
affected by local scattering material, Doppler factors of $\sim$ 1000
are not required to explain this source. The discovery of a second
source whose properties are well modeled by interstellar
scintillation strengthens the argument for this as the cause for
much of the variations seen in intra-day variables (IDV).

\end{abstract}
\notetoeditor{If possible fig.2 should be set across two columns, not
vertically in one column}

\section{Introduction}
Variability in extragalactic radio sources is used to probe the
smallest regions of these sources. Many flat-spectrum quasars and
BLLacs are intrinsically variable (Aller et al, 1985), typically on
timescales of weeks to years. A class of rapid variables -- the
intraday variables (IDV, Witzel et al, 1986) -- have recently
attracted attention. If the observed variations in these sources are
intrinsic, then causality implies brightness temperatures far in excess
of the Compton Catastrophe limit (when all the energy in synchrotron
radiating electrons should be rapidly converted to X-ray photon
energy.) To resolve this problem it has been suggested that the
variations are not intrinsic, but due to a propagation effect known as
scintillation, but the question remains unresolved (see Wagner \&
Witzel, 1995).

Interstellar scintillation is demonstrated in pulsars at radio
frequencies (Rickett, 1977), and is probably responsible
for low frequency variability in flat spectrum radio sources (Rickett
et al, 1984). The discovery of dramatic variability at GHz frequencies
in PKS\,0405-385 was explained by Kedziora-Chudczer et al. (1997) as a
scintillation effect.

In this {\it Letter} we report on the discovery of a source whose variations
are also best explained as a scintillation effect.  J\,1819+3845 is a
faint $\sim$\,21\,mag quasar at 18h19d26.55s +38d45m01.8s (J2000).  An
optical spectrum obtained at the INT on 1999 May 10, showed broad emission
lines at z=0.54.  The radio source has no detected extended emission,
and has a rising radio spectrum (fig.1) with $\alpha \sim\,-0.5$ (we
use S$\propto\nu^{-\alpha}$).

Throughout the paper we assume a q$_o$=0.5, $\Lambda$=0 cosmology with
H$_o$ = 65\,kms$^{-1}$Mpc$^{-1}$. 

\begin{figure}
\psfig{file=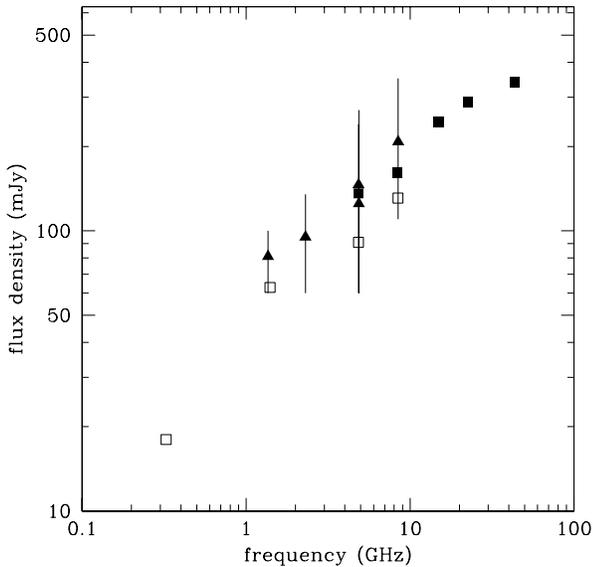,height=8cm}
\caption{Radio spectrum of J\,1819+3845. The open symbols are survey data
(WENSS, NVSS, GB6, CLASS). The filled symbols are new observations:
squares VLA (Mar 98); triangles mean WSRT flux densities. Error bars
for all observations fall within the symbols. The vertical lines
indicate the flux density range in the WSRT obervations reported here.}
\end{figure}

\section{Observations}
Observations of J\,1819+3845, as a member of a sample of about 50 CLASS
(Myers et al., 1999) sources, started in January 1999 with two
flux density measurements separated by 4 days. These indicated a
factor 2 variation in flux density.  A further 10 observations, at a
variety of time intervals, were made in late March 1999. These showed
the same flux density range and indicated that the variability
timescale was at most an hour. A long 96 hour campaign was then
conducted from May 13 UT 0300 until May 17 0300 to study J\,1819+3845, as
well as other sources.  Here we report only on the observations of
J\,1819+3845. 

The observations were obtained with the new WSRT Multi Frequency
Frontends (MFFEs) (see \url{http://www.nfra.nl/wsrt}).  Data were
taken at 1.4, 2.3, 4.9 and 8.4\,GHz using a backend with 8 contiguous
bands, each of 10 MHz bandwidth.  Standard calibrators were observed
for about 5 min once every hour for amplitude and phase calibration.
The Baars et al. (1977) flux scale was used for 3C286 and
all other sources are tied to 3C286.

In the 96\,hour campaign we observed J\,1819+3845 for two 12\,hour
periods with two sub-arrays each of 6 telescopes which had frontends
tuned to a different frequency and a total bandwidth of 4$\times$10\,MHz.  On
May 13/14 we observed at 4.9/1.4\,GHz, the following night we observed
at 4.9/8.4\,GHz.  The time resolution was 30\,seconds for the
split-array observations, and 10\,seconds for all other data.  With a
source flux density of typically 100\,mJy, noise was clearly
insignificant (see table 1).

The normalized rms fluctuations $S_{rms}/<S>$ (modulation index, $m$)
are calculated for each 12\,hour run, and for the intermittent
observations over the full 96\,hours, and tabulated in table 1. The
error is obtained by calculating $m$ in the first and last half of the
run independently. The characteristic timescale of the fluctuations is
calculated by use of the structure function D$^1$ (see e.g. Simonetti,
Heeschen \& Cordes, 1985).  The characteristic timescale is taken as
twice the time for D$^1$ to reach $(1-e^{-1})$D$^1_{max}$=
$(1-e^{-1})2m^2$.  This corresponds to the full-width at $1/e$ of the
auto-correlation function.  The structure function has been corrected
for noise bias using 2$\sigma_{\delta f}^2$, in the manner of
Simonetti et al (1985). 

\begin{figure*}
(a)
\psfig{file=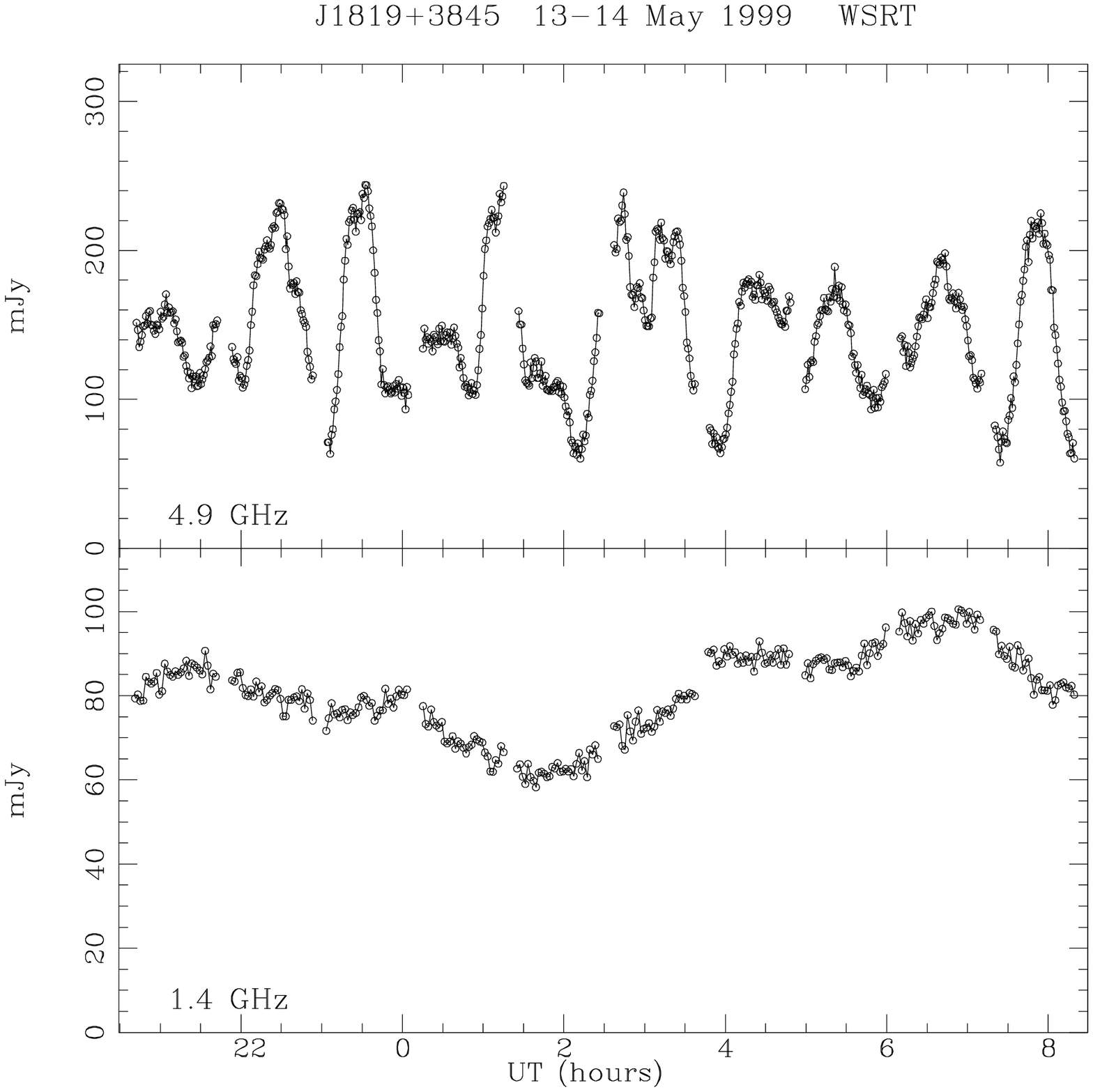,height=8cm}
(b)
\psfig{file=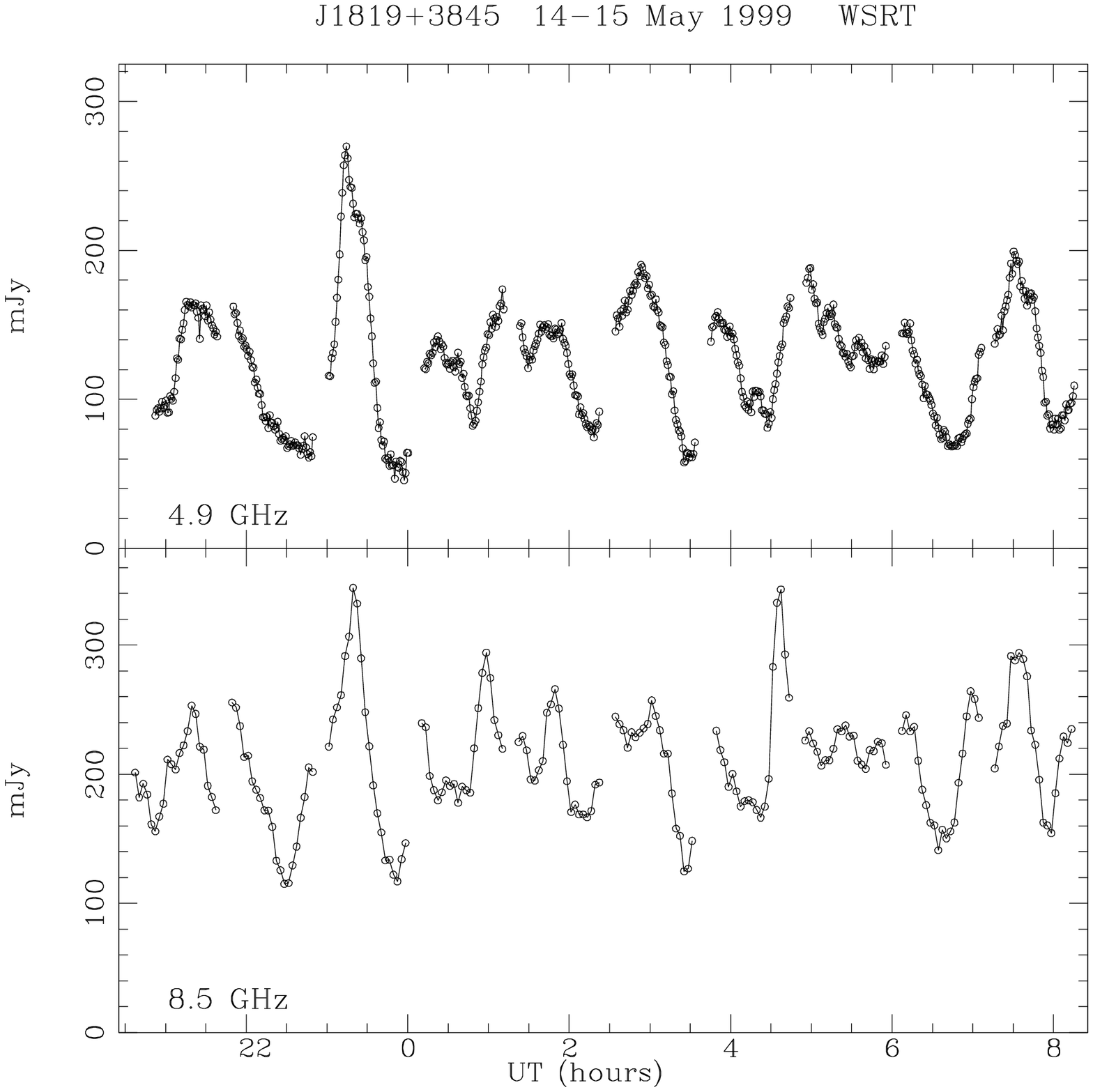,height=8cm}
\caption{The light curves of J\,1819+3845 at (a) 12 hour WSRT
split-array observations at 1.4 and 4.9\,GHz. (b) 4.9 and 8.5\,GHz.}
\label{fig:light}
\end{figure*}

\newpage

\section{Source size and structure}

If the variations are intrinsic then the source size must be of the
order of a light-hour. At a redshift 0.54, this corresponds to
nanoarcseconds. A source this small will necessarily scintillate, so
we only consider this as the cause of the short timescale
variations.

Weak scintillation occurs at high frequencies where the diffractive
scale of the turbulence is smaller than the Fresnel scale. At lower
frequencies two different branches of scintillation develop:
refractive and diffractive (see Narayan 1992 for a
clear exposition). The transition between the regimes of
diffractive/refractive scintillation and weak scintillation occurs at
a critical frequency, when the refractive, diffractive and Fresnel
scales are equal. In order to show large modulations at this
frequency, the apparent source size must be smaller than, or
comparable to, the Fresnel scale i.e.  $\sqrt{\lambda/2 \pi L}$,
where L is the distance to the `equivalent screen'.  At the critical
frequency there is a very characteristic peak of the modulation index
of broad-band scintillations.  The intensity of the fluctuations in
J\,1819+3845 peaks around 5\,GHz and the timescales of the modulations
get much longer below this frequency, similarly to PKS\,0405-385
(Kedziora-Chudczer et al 1997), providing strong evidence for
scintillation as the cause of the variations (table 1, fig.3).

The timescale at the critical frequency is $\approx \sqrt{\lambda
L}/v$. The timescales of the variations in J\,1819+3845 at (the close to
critical) frequency 5\,GHz occur much too rapidly for the equivalent
screen to be at a distance of $\sim$1\,kpc, as in the model of Walker
(1998). For a relative screen--earth velocity of 50\,km/s (typical
earth rotation speed around the sun) this predicts a FWHM timescale of
$\sim$ 6\,hr.

\begin{table*}
\caption{Observed quantities}
\begin{tabular}{lccrrrrrrrr}
\tableline
Freq & date & MJD&rms/60s&$<S>$ &m & $\sigma_m$ &m$^\prime$& t & $\sigma_t$ &
$\sigma_{\delta f}$\\
GHz   &1999& (start)  &mJy  &mJy&&&&mins&&\\
\tableline
1.340--1.380&May 13-14&51311 &1.4&81  &0.13 & 0.02&0.24& 212&12& 0.03\\
1.340--1.420&May 13-17&51311 &1.0&87  & 0.12& 0.02&0.21&  --&-- & --\\
2.210--2.290&May 13-17&51311 &1.0&95  & 0.24& 0.04&0.44&  --&-- & --\\
4.834--4.874&May 13-14&51311 &2.8&146 & 0.29& 0.01&0.52& 34 &4 & 0.17\\
4.834--4.914&May 14-15&51312 &2.0&125 & 0.32& 0.03&0.58& 32 &2 & 0.05\\
8.450--8.490&May 14-15&51312 &2.5&208 & 0.21& 0.01&0.38& 30 &6 & 0.10\\
\tableline
\end{tabular}
\small{

m : modulation index, or the normalised rms fluctuations in intensity\\
m$^\prime$ :  modulation index with 55\% of the flux density in the
scintillating component\\
t : timescale for the variations. Error on
timescale  $\sigma_t$ determined by omitting $\sigma_{\delta f}$ correction}
\end{table*}

Using the measured timescale of the variations and the critical
frequency, we can solve for the scattering measure (or equivalently
C$_N^2$=10$^4$\,C$_{-4}$; Armstrong et al, 1981), the radius of the
scattering disk ($\Theta$), and the distance to the equivalent screen
($L$), if we assume the relative transverse velocity of source--screen
projected onto the earth ($v$). The scattering disk $\Theta$
corresponds to maximum source radius for full scintillation. The
apparent source velocity projected onto a screen within our Galaxy is
less than a few km/s, even for apparent source speeds of 10$c$,
therefore the velocity is expected to be dominated by the motion of
the earth. Using the analytic approximation of Blandford \& Narayan
(1985), for a Kolmogorov spectrum of irregularities in a thin screen,
we solve:
$$\Theta_{\mu as}=0.012 v_{kms^{-1}} t_{hrs} L^{-1}_{kpc}$$
$$C_{-4}=22.4\Theta^{5/3}_{\mu as}\lambda_{cm}^{-11/3}L^{-1}_{kpc}$$
$$\lambda^{crit}_{cm}=2.77 C_{-4}^{-6/17} L^{-11/17}$$ Taking the
critical wavelength $\lambda^{crit}$=$6\pm1.5$cm, $v$=50\,km\,s$^{-1}$
and t=0.5\,hrs we obtain\\ C$_N^2$=0.17$\pm0.05$m$^{-20/3}$, an
effective distance to the scattering screen of 19$^{-4}_{+6}$\,pc, and
a source radius of $16\pm4\mu$arcsec. We note that this theory is only
strictly valid in the weak scattering regime, but that the source size
and distance so derived agrees well with the more simple derivation
above.

The 5\,GHz light curve is repeatedly seen to drop to values as low as
60\,mJy. The consistency of the `base-level' provides evidence for a
non-scintillating (larger) component and provides a strong constraint
of $\eqslantless$\,60\,mJy in this component (and implies
$\eqslantgtr$\,75\,mJy in the scintillating component, taking 135\,mJy
as the average flux density). Fig.~3 shows the observed $m$, and the
curves obtained with the maximum possible flux density in the extended
material, and the scintillating component as smaller than the
scattering disk.

We considered the case that all the source flux density is contained
in a region somewhat larger than $\Theta$, allowing the source size to
vary as a negative power of frequency.  In order to reproduce the
fluctuation indices observed, the source cannot be more than $\sim$10
times the Fresnel scale at the critical frequency. However, the very
rapid variations at 5\,GHz then require an extremely strong scattering
screen (C$_N^2\sim 10^4$m$^{-20/3}$) less than a parsec
distant. As there is no evidence for such material (e.g. from pulsar
studies), we do not subscribe to such a model.

\begin{figure}
\psfig{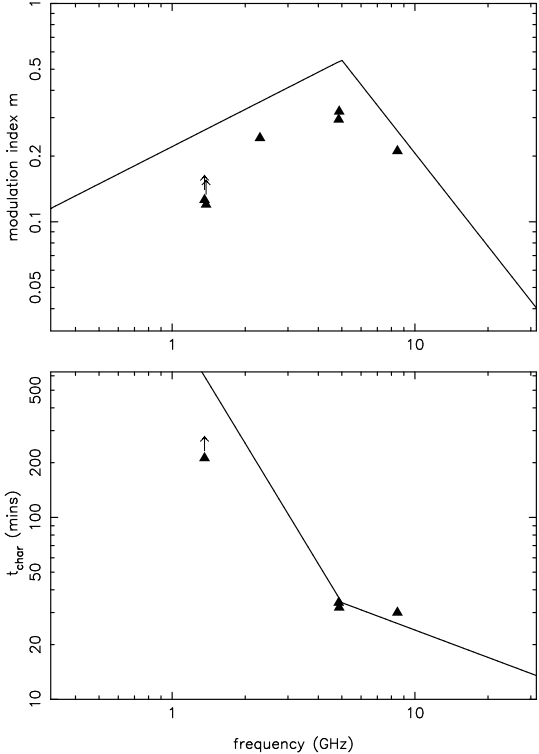}
\caption{Observed modulation index and time-scales overlaid on
thin-scattering screen model predictions, for a scintillating
component plus a non-scintillating region. There is 55\% of the total
flux density in the scintillating component at all frequencies. The
critical frequency taken as 5\,GHz where we take m=1 for a point
source, and the predicted timescale is normalised to the timescale at
5\,GHz.}

\end{figure}
We have searched for (even faster) diffractive scintillation in
observations with a bandwidth of 5\,MHz at 1.38\,GHz, but found none
(m$_{diff}<0.1$). However, interpretation of this as a source size is
complicated by the unknown flux density of the non-scintillating
component and the possibility that the decorrelation bandwidth may be
$<$5MHz. Further, for sources with 
m$_{diff}<1$, the predicted diffractive timescale
increases and becomes closer to the refractive one.

\section{Discussion}

J\,1819+3845 must have at least 75\,mJy within a radius 16\,$\mu$arcsecs
at 5\,GHz, if 5\,GHz is the critical frequency.  This results in an
observed brightness temperature T$_b$ $>$ 5.10$^{12}$K.  A higher
critical frequency results in a higher brightness temperature for a
flat or rising spectrum of the scintillating component. 

The scattering occurs closer to the observer than is predicted by a
uniform weighting of the TC93 model.  A screen location at
$\sim$25\,pc is, however, in agreement with the model of the Local
Bubble by Bhat, Gupta \& Rao (1997), to which the authors attributed
enhanced scattering of pulsars. If the scattering occurs further away,
a smaller source size and higher T$_b$ is required, although, as
outlined above, this is difficult to reconcile with the data. We also
point out that the contribution to the scattering of a source of
finite size seen through an extended medium will be greatest where the
Fresnel scale is equal to the angular size of the source. This effect
will decrease the distance of the effective scattering screen.

If we assume that the rising spectrum is caused by synchrotron
self-absorption, from an isotropic particle distribution producing
optically thin emission above $\nu_{obs}$, we calculate that a Doppler
factors (\dopp~) of 12--15 are required to explain the apparent
T$_b$. (We use the formula of Readhead (1994) and consider both the
high frequency cut-off to be 10$\times \nu_{obs}/$\dopp~ and the
canonical cut-off of 100GHz, with a particle energy index of 2,
$\nu_{obs}$=43\,GHz).  An obvious candidate for such emission is an
optically thick component of an extremely fast, sub-parsec jet.  This
should be moving at $\sim$ 1\,mas per year: readily detectable by VLBI.

However, such an object is suffering large Compton losses, and will
survive much less than a year, without input of fresh energy. In the
observer's frame, the half-life of the electrons emitting at the peak
frequency is only a couple of light weeks, due to time dilation. Such
effects cannot be avoided by a faster moving blob, as the
effect of time dilation increases faster than the increase in energy
loss timescale.  This energy loss is apparently not seen: average
5\,GHz flux densities in January, March and May 1999 are 123, 135, and
135\,mJy respectively. J\,1819+3845 is also apparently stable in the
long term: the available data at all frequencies spans over 10 years,
and show no changes greater than a factor of 2 (fig.1).

We cannot yet rule out that the source may be `transient' and
radiating above the Compton limit for a period no longer than the
light-crossing time of the source, after a single injection of
energy. However, the remarkable fact that the source showed the same
modulation index (flux density range of a factor 3) over a period of
at least 2, probably 4, months suggests to us that the emission
comes from a steady region in space through which emitting particles
flow.  We suggest two broad pictures, both of which require a
continuous internal source of energy:

(a) a `cauldron' near the AGN, of which we see the $\tau \approx 1$
surface. The particles are continuously regenerated near the black
hole and lose energy exceedingly rapidly.  

(b) a relativistic wind of particles or jet in which the particles
stream out. Under these conditions inverse Compton losses
may be seriously reduced (e.g. Woltjer 1966)

PKS\,0405-385 is also an inverted spectrum quasar (Kedziora-Chudczer et
al 1997) as is PKS\,1741-038 whose variations have also been interpreted as
scintillation (Hjellming \& Narayan 1986). We suggest
that the unusual spectral shape is an important feature.  The overall
shape and relative stability of the radio spectrum (0.3 -- 43\,GHz) is also reminiscent of that of M81 (de Bruyn et al, 1976;
Reuter and Lesch, 1996) as well as SgrA* (Falcke et al, 1998).

\section{Conclusions}
The combination of exceedingly large modulations in the received flux
density and its very short variability timescales make the quasar
J\,1819+3845 the most extremely variable source known in the radio
sky. At 5\,GHz its modulation timescale is half an hour, and the peak
to peak variations are frequently 350\%. We interpret the variations
as due to interstellar scintillation, and derive an angular diameter
at 5\,GHz of $<$32\,$\mu$arcsecs.

The apparent brightness temperature is $>$\,5.10$^{12}$K. Unlike the
`variability' T$_b$ observed in IDVs, this is (almost) a direct
measurement.  The source J\,1819+3845 is at most five light-months in
diameter. Considering the object as a plasma blob, isotropically
emitting in its rest frame, requires bulk relativistic motion with
\dopp$\sim$15 to explain the observed brightness
temperature. However, the inverse Compton energy loss timescales for
such an object are a few weeks, but we have seen similar mean flux
densities and scintillation over 4 months. 
We favour models in which energy is supplied continuously to the emitting
region, and we are observing a `constant surface' of the source
through which particles flow.

A very high T$_b$ (10$^{14}$K) was derived for another quasar
PKS\,0405-385, whose variability was also interpreted as an effect of
scintillation (Kedziora-Chudczer et al 1997). Such a high brightness
temperature would not be needed if the effective scattering screen was
placed closer to the observer than assumed by Kedziora-Chudczer et
al. The interpretation of IDV variations as intrinsic implies source
sizes which are often small enough for scintillation to play a role
(Rickett et al 1995).  Good evidence for scintillation in two sources
strengthens the case that IDV variations are predominantly caused by
scintillation effects.

\acknowledgments

We are grateful to Rene Vermeulen for allocating generous amounts of
WSRT 'backup' and 'filler' time.  We thank Gie Han Tan and his
MFFE-team for the timely completion of wonderful series of
frequency-agile receivers for the WSRT and Hans van Someren Greve for
his last-minute software efforts to enable split-array multi-frequency
operation.

The WSRT is operated by the Netherlands Foundation for Research in
Astronomy (NFRA/ ASTRON) with financial support by the Netherlands
Organization for Scientific Research (NWO). The VLA is operated by
NRAO which is a facility of the National Science Foundation, operated
under cooperative agreement by Associated Universities, Inc.
This research was supported by the European Commission, 
TMR Programmme, Research Network Contract ERBFMRXCT96-0034 'CERES'.

\end{document}